# Nanoporous gold leaves: plasmonic behavior in the visible and mid-infrared spectral regions

Denis Garoli[1*], Gianluca Ruffato[2], Pierfrancesco Zilio[1], Eugenio Calandrini[1], Francesco De Angelis[1], Filippo Romanato[2] and Sandro Cattarin[3]

[1]*Istituto Italiano di Tecnologia, Via Morego 16, 16136 Genova, Italy*
[2] *Department of Physics and Astronomy, University of Padova, Via Marzolo 8, 35131 Padova Italy*
[3]*Institute for Energetics and Interphases (CNR-IENI), Corso Stati Uniti 4, 35127 Padova Italy*
[*]*denis.garoli@iit.it*

**Abstract:** A robust and reproducible preparation of self-standing nanoporous gold leaves (NPGL) is presented, with optical characterization and plasmonic behaviour analysis. Nanoporous gold (NPG) layers are typically prepared as thin films on a bulk substrate. Here we present an alternative approach consisting in the preparation of NPGL in the form of a self-standing film. This solution leads to a perfectly symmetric configuration where the metal is immersed in a homogeneous medium and in addition can support the propagation of symmetric and antisymmetric plasmonic modes. With respect to bulk gold, NPG shows metallic behaviour at higher wavelengths, suggesting possible plasmonic applications in the near / medium infrared range. In this work the plasmonic properties in the wide wavelength range from the ultraviolet up to the mid-infrared range have been investigated.

## 1. Introduction

Porous gold films have recently attracted increasing interest due to the unique properties related to their very high specific surface area. This particular material finds applications in many fields from electrochemistry [1] to nanofluidic [2], solar cell [3] and enhanced spectroscopy [1, 4]. Among the different fields of application, nanoporous gold (NPG) demonstrates also intriguing properties as plasmonic material. There is still a great desire to develop materials with optimized photonic properties taking advantage of electromagnetic resonances and concomitant enhancement of the electromagnetic near-field owing to surface plasmons. In the continuous search for new materials capable of better performance, NPG offers an extensive surface where the analytes can bind, leading to an increased sensitivity [5-8]. During the last years we have extensively studied the plasmonic properties of NPG [6-10], with an eye to sensing applications. The material, prepared on a bulk substrate, has been properly patterned with periodic structures in order to support the excitation of propagating plasmonic modes. Both the architectures exploiting reflectance and transmittance have been tested. In all cases NPG layers showed important advantages for sensing applications with respect to bulk gold layers patterned with the same methods and criteria [6-8], thanks to the enhanced surface-to-volume ratio and the concomitant excitation of localized surface plasmon modes. In this paper we aim at extending those experimental verifications by considering the optical properties and testing the sensing capabilities of a self-standing NPG film.

Several methods for the synthesis of porous gold films are reported in the literature, among others: chemical or electrochemical de-alloying [11-13], electrodeposition in assemblies of templating agents [14-16] or in porous membranes [17], template free electrodeposition [18-19], casting of a suspension of gold nano-particles on a gold substrate [20]. The optical characterization of NPG thin films has been presented by several authors [7, 21, 22]. Conversely, self-standing NPG leaves have been prepared by relatively few authors [23-28], and an exhaustive optical characterization in the spectral range between 200 and 900 nm has been performed only recently by Detsi et al. [26]. The structure of self-standing leaves provides unique conditions for the optical characterization of NPG since the absence of a supporting metal substrate avoids interface effects and eliminates ambiguities in the analysis of the collected data. Moreover, looking at the related plasmonic behaviour, a self-standing metallic layer represents a perfect symmetric configuration (being at the same time easily accessible to analytes, both gases and liquids) and it can be exploited for the excitation of different plasmonic modes in a wide spectral range.

In the present work, self-standing NPG leaves are prepared by chemical dealloying of a gold-silver alloy thin film and investigated by spectroscopic ellipsometry and infrared spectroscopy in order to extend the characterization of the material optical behavior in the spectral range from the ultraviolet (300 nm) up to the mid-infrared wavelengths (15 μm). The resonant pattern is analyzed and explained by means of a Drude-Lorentz model/analysis including inter- and intra-band adsorptions and localized plasmon excitation. Finally, a comparison of the optical response before and after the functionalization with a self-assembling monolayer of dodecanethiol is performed in the optical and near-infrared spectral range. The dispersion curve is numerically calculated in the visible and infrared spectral region and the plasmonic modes there supported are considered and discussed.

## 2. Experimental

*2.1 NPG Leaf preparation*

Samples of 6 carat gold leaf were purchased from Wasner Blattgold GmbH (Schwabach, Germany). According to supplier specifications, the mass composition is Ag 75%, Au 25%, corresponding to an atomic composition close to $Ag_{84}Au_{16}$, and the average thickness is about 140 nm. Previous literature investigations, performed on the same material [25, 29] report an atomic composition (evaluated by energy dispersive x-ray, EDS) $Ag_{80}Au_{20}$ and an estimated thickness of 120 nm.

Sample preparation is schematically illustrated in Fig. 1. Prior to the etching procedure, a Ti rectangular sheet 1.5x10 cm was cut from a thin pure Ti foil (0.125 mm thickness, Goodfellow); several small holes (0.7 mm diameter) were made with a drill on the Ti sheet (Fig. 1a). After careful polishing to clean the surface and eliminate burrs, a thin Au layer (100 nm) was electrochemically deposited from a commercial bath (TECHNI GOLD 25 ES) onto the Ti sheet, to be used for subsequent handling of the nanoporous gold leaf (NPGL) in the etching solution. The gold layer aimed at avoiding chemical attack and release of oxidised Ti species following contact with nitric acid. The commercial 6 carat gold leaf was laid down with care on the surface of a concentrated nitric acid solution (65%, Aldrich), Fig. 1b; the colour of the floating leaf changed rapidly from the original silver colour to a copper tone, indicating formation of nanoporous gold. The leaf was left etching for 30 min, then carefully transferred by means of the gold coated Ti sheet on the surface of distilled water on which it was left rinsing for further 30 min. Thereafter, the Ti sheet was immersed in the solution and extracted slowly so as to get the NPGL laying on its surface (Fig. 1c). In correspondence of the holes a self-standing NPGL was present and could be investigated (Fig. 1d). Ellipsometric measurements were performed on a NPGL deposited on a glass plate substrate.

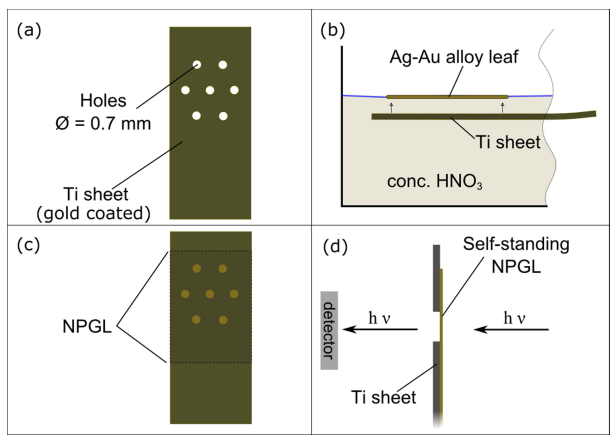

Fig. 1. Scheme (not to scale) illustrating sample preparation: a) (gold coated) Ti sheet with small holes; b) a 6 carat gold leaf is deposited on the concentrated $HNO_3$ solution and left etching for 30 min, then transferred, rinsed and extracted by means of the Ti sheet; c) the dry NPGL finally lays on the Ti support; d) a self-standing NPGL is obtained in correspondence of the holes.

*2.1 Film characterization*

The morphology and the composition of the prepared self-standing leaves were investigated by means of scanning electron microscopy (SEM) and EDS (by using a FEI HeliosNanolab650 dual beam system equipped with Oxford Instruments® EDS system). Fourier transform infrared spectroscopy (FTIR) was used to measure the material dielectric constants in the spectral range 1900 – 15000 nm (0.08 - 0.65 eV). In particular, analyses (both in transmission and in reflection mode) of self-standing films were carried out with a commercial micro-FTIR (ThermoFisher iS50), equipped with a cassegrain condenser lens which mimics a 15X, NA=0.58 objective (200 scans with a spectral resolution of 0.005 cm$^{-1}$). The absolute transmission (T) and reflectivity (R) of the NPGL were measured using the empty channel and a gold mirror as reference, respectively, collecting the light in different positions of the sample to check the reproducibility of the results. The unknown optical constants of the NPGL are extracted from the fitting of T and R to a multilayer model composed of a NPGL layer sandwiched between two air layers. The thickness of the NPGL layer was determined during the morphology analysis. Within the Drude model for the dielectric function ε, the response of the free carriers of NPGL is then entirely determined by the screened plasma frequency $\omega_p$ and the scattering rate $\Gamma$ according to the following formula

$$\varepsilon(\omega) = \varepsilon_\infty - \frac{\omega_p^2}{\omega^2 + i\omega\Gamma} \qquad (1)$$

Spectroscopic ellipsometry (SE) was performed by a VASE Spectroscopic Ellipsometer (J.A. Woollam) with angular and wavelength resolutions of 0.005° and 0.3 nm, respectively. The goniometer controlled optical bench was set for three different angles of incidence on the sample (50°, 60°, 70°) and the ellipsometric data were recorded in the rotating polarizer analysis setup (RAE) [30] in the range 300-2400 nm (4.13 – 0.52 eV), step size 10 nm. The output of ellipsometric analysis is the ratio ρ of the p-polarized and s-polarized complex Fresnel reflection coefficients $r_{pp}$ and $r_{ss}$, expressed in terms of the ellipsometric angles ψ and Δ:

$$\rho = \frac{r_{pp}}{r_{ss}} = \tan\psi e^{i\Delta} \qquad (2)$$

Afterwards data were analyzed with WVASE32 software (J.A. Woollam), provided by the manufacturer. Samples are usually modeled as a stack of n layers, depending on sample complexity, each one characterized by its complex dielectric function and its effective thickness. Thus experimental data are fitted until the best agreement with the model is achieved and an estimation of the optical constants and thickness of each layer is provided. Ellipsometric measurements were performed on the nanoporous gold leaf lying on the glass substrate. A comparative ellipsometric analysis of bare glass substrate and of the NPGL on glass sample allows extracting the complex permittivity $\varepsilon_1 + i\varepsilon_2$ and thus the complex refractive index $n + ik$ of the nanoporous gold film.

Transmission spectra were acquired in normal incidence with a wavelength spectroscopic resolution of 0.3 nm, using the monochromatized 75W Xe lamp of the ellipsometer setup. In this case the self-standing NPGL was measured.

In order to test the biosensing capabilities of the self-standing leaves, ellipsometric measurements were collected before and after the functionalization of gold ligaments with a suitable molecule in air environment. A self-assembled monolayer of dodecanethiol ($C_{12}H_{25}SH$) was deposited on the gold surfaces at room temperature. The sample was gently pre-cleaned in a basic peroxide solution (5:1:1 double distilled $H_2O$, 30% $H_2O_2$ and 25% $NH_4OH$) for 10 minutes, rinsed in distilled water and dried under $N_2$ flux. The cleaned sample was submerged in 6-mM solution of dodecanethiol in ethanol for about 48 hours and therefore rinsed thoroughly with ethanol for at least 5 minute, followed by drying under a gentle nitrogen stream. The spontaneous assembly of the molecules is known to form a densely packed and highly oriented structure on a metallic surface [31].

## 3. Results and Discussion

### 3.1 Morphology and composition

Figure 2 reports SEM micrographs of the NPGL. The microstructure consists of a fully connected network of gold ligaments and pores. EDS analysis indicates that the residual silver content is about 3%, in substantial agreement with previous reports of exhaustive etching of very thin foils [25] and massive alloy samples [32, 33]. Moreover, from our EDS analysis a slight contamination of Titanium (below 3%, presumably $TiO_2$) was detected. This is reasonably due to the acid attack on the titanium foil used as support for the thin film obtained after the dealloying. Evaluation of specific surface area (surface area per unit mass) was not attempted due to experimental difficulties, but the broad range 2 to 10 m²g⁻¹, estimated in previous investigations of porous Au films [34], should be a correct interval of values.

The dealloying process entails a significant volume contraction, estimated at about 20% for bulky samples [29] and thin films [35] of composition similar to ours. Considering that Au and Ag have nearly identical lattice parameters [35], in the absence of shrinking the occupied volume fraction in porous gold would correspond to the fraction of Au atoms, assuming total removal of Ag atoms. Taking into account the volume shrinking (to 80% of the initial), and neglecting the small fraction of residual Ag, the effective density of the NPGL relative to the density of bulk gold may be estimated between $0.16/0.80 \cong 0.20$ (using the formula $Ag_{84}Au_{16}$ obtained from the supplier) and $0.20/0.80 \cong 0.25$ (using the formula $Ag_{80}Au_{20}$ estimated in Ref. 29).

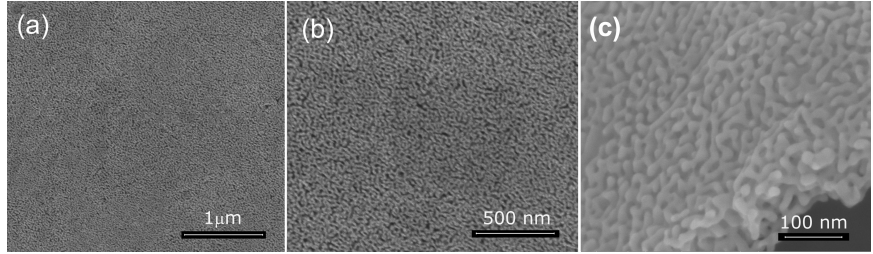

Fig. 2. SEM micrographs of a self-standing NPGL sample: a) and b) top view showing surface morphology; c) tilted view showing section and inner morphology.

### 3.2 Spectroscopic ellipsometry

Figure 3 reports the experimental ellipsometric data and the obtained dielectric constants. In the range from near-UV to near-IR, the permittivity is well described with a Lorentz-Drude model:

$$\varepsilon(\omega) = \varepsilon_\infty + \varepsilon_{UV}(\omega) + \varepsilon_D(\omega) + \varepsilon_{IR}(\omega) \qquad (3)$$

$$\varepsilon(\omega) = \varepsilon_\infty - \frac{A_{UV}}{\omega^2 - \omega_{UV}^2 + i\omega\omega_{\tau,UV}} - \frac{\omega_p^2}{\omega^2 + i\omega\omega_\tau} - \frac{A_{IR}}{\omega^2 - \omega_{IR}^2 + i\omega\omega_{\tau,IR}} \qquad (4)$$

where $\varepsilon_\infty$ takes into account the constant contribution to polarization due to $d$ band electrons close to the Fermi surface. $\varepsilon_{UV}(\omega)$ is a Lorentz oscillator that describes the $3d$ energy band-to-Fermi Level interband transition centered at a frequency $\omega_{UV}$ in the UV range with a band-width $\omega_{\tau,UV}$. The Drude contribution of the free s-electrons is described by $\varepsilon_D(\omega)$, with plasma frequency $\omega_p$ and relaxation time $\tau = 2\pi/\omega_{\tau,D}$. The Lorentz contribution $\varepsilon_{IR}(\omega)$, with amplitude $A_{IR}$, centered at a frequency $\omega_{IR}$ and band-width $\omega_{\tau,IR}$, is necessary to properly model the behavior of the dielectric response in the near IR range. This extra term enables excellent fits to dielectric constant of nanoporous films [36] and is associated to the excitation of localized surface plasmons [36,37].

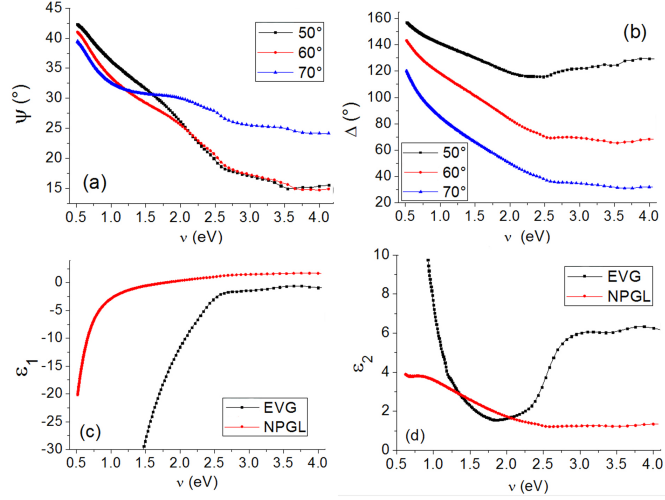

Fig. 3. (a,b) Ellipsometric angles ψ (a) and Δ (b) from spectroscopic ellipsometry in the range 400-2400 nm, step 10 nm, incidence angles 50°-60°-70°, of nanoporous gold leaf lying over a glass substrate. (c,d) Dielectric permittivity of nanoporous gold leaf (NPGL) (solid line): real (c) and imaginary (d) parts. Comparison with evaporated bulk gold (EVG) (dashed line)..

**Table 1. Fitting parameters from the analysis of nanoporous gold and evaporated gold layers with oscillators model of Eq. 4.**

|  | $\varepsilon_\infty$ | $A_{UV}$ (eV$^2$) | $\omega_{UV}$ (eV) | $\omega_{\tau UV}$ (eV) | $A_{IR}$ (eV$^2$) | $\omega_{IR}$ (eV) | $\omega_{\tau IR}$ (eV) | $\omega_p$ (eV) | $\omega_\tau$ (eV) | $N$ $10^{21}$cm$^{-3}$ | $\tau$ (fs) | Au% | $p$ |
|---|---|---|---|---|---|---|---|---|---|---|---|---|---|
| NPGL | 1.81 | 21.01 | 4.72 | 3.36 | 11.20 | 1.32 | 2.79 | 2.58 | 0.66 | 4.84 | 6.3 | 22 | 0.24 |
| EVG | 5.92 | 86.6 | 4.71 | 3.15 | -- | -- | -- | 8.36 | 0.41 | 50.81 | 10.1 | 100 | -- |

Permittivity values calculated from ellipsometric analysis have been fitted with the oscillator model and an estimation of the fitting parameters has been obtained. Results are collected in Table 1. The resulting estimated thickness of the NPGL is 116.7 ± 4.1 nm.

The intraband absorption term in the UV range is weaker than in bulk gold, however its position around 4.7 eV is preserved since it is related to gold atomic and band structure which results weakly perturbed by the reorganization of gold atoms into ligaments with average size of 15-20 nm. The reorganization of gold atoms affects the relaxation time $\tau$ of free carriers in the Drude term which is shorter than for bulk gold, since it is related to the scattering processes in the material. The density $N$ of free carrier can be estimated from plasma frequency [38] $\omega_p$:

$$\omega_p = \sqrt{\frac{Ne^2}{m\varepsilon_0}} \qquad (5)$$

where $e$ and $m$ are respectively electron charge and mass, $\varepsilon_0$ is the void dielectric permittivity. As expected, free-charge density in nanoporous gold results lower than the value $N_{bulk}$ = 50.81·10$^{21}$ cm$^{-3}$ for bulk gold, but the greater void fraction is not the only responsible. If the porous gold density is assumed to be around 25% that of bulk gold, the expected free-electron density should be around the value 0.25 · 50.81·10$^{21}$ cm$^{-3}$ ≅ 12.7·10$^{21}$ cm$^{-3}$, quite greater than the experimental value $N_{NPGL}$ = 4.84·10$^{21}$ cm$^{-3}$. It is clear that ligament size and shape affect the effective free-electron density and have an effect on the metallic response. In order to properly describe this phenomenon, a Bruggeman Effective Medium Approximation (BEMA) [39] has been employed to model the effective dielectric permittivity $\varepsilon_{eff}$ of nanoporous layer by assuming the medium as a system of gold inclusions into a void matrix:

$$(f_{Au})\frac{\varepsilon_{eff} - \varepsilon_{Au}}{\varepsilon_{Au} + p(\varepsilon_{eff} - \varepsilon_{Au})} + (1 - f_{Au})\frac{\varepsilon_{eff} - 1}{1 + p(\varepsilon_{eff} - 1)} = 0 \quad (6)$$

where $f_{Au}$ is the gold fraction, $p$ is the effective depolarization factor of gold inclusions that takes into account the geometric effect of gold ligament size and shape on electron-plasma oscillations. By modelling the nanoporous gold layer with a BEMA film with thickness 116 nm, we get a gold fraction estimation $f_{Au} = 0.22 \pm 0.01$ and a depolarization factor $p = 0.24 \pm 0.02$. The gold percentage of 22% thus estimated is about halfway the range of relative density 0.20-0.25 estimated at the beginning of this section.

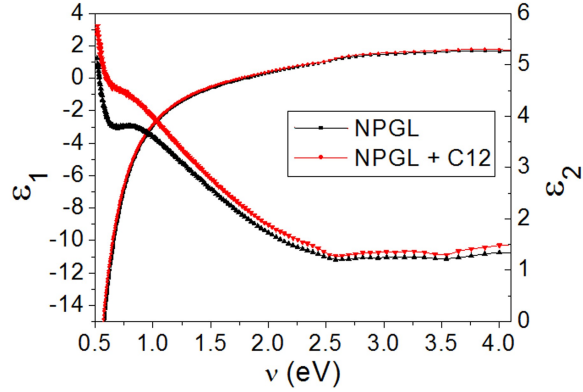

Fig. 4. Dielectric permittivity real and imaginary part before (dark line) and after (red line) functionalization with C12.

Table 2. Plasmon resonance in nanoporous gold before and after functionalization with C12. Fitting parameters from analysis of dielectric permittivity with oscillators model of eq. 4 and resulting wavelength shift $\Delta\lambda$ of the resonance peak in the near-infrared range.

|  | $A_{IR}$ | $\omega_{IR}$ (eV) | $\omega_{\tau,IR}$ (eV) |
|---|---|---|---|
| NPGL | 11.200 ± 0.103 | 1.324 ± 0.003 | 2.790 ± 0.023 |
| NPGL + C12 | 14.411 ± 0.139 | 1.269 ± 0.003 | 3.159 ± 0.028 |
| $\Delta\lambda$ (nm) | **41.0 ± 4.4** | | |

The previous ellipsometric analysis has been repeated after sample functionalization of gold ligaments with a self-assembling dodecanethiol coating. As widely described and exploited in literature [40], surface plasmon modes, either localized or propagating, are extremely sensitive to the physical and chemical conditions of the supporting surface, as a consequence of the confinement of the electromagnetic field close to the interface. A change in the effective refractive index experienced by these modes, for instance due to a binding of molecules to the supporting metal, causes a variation in the propagation constant and therefore a change in the coupling conditions, which reflects upon a shift of the resonance wavelengths. Therefore a shift of the resonance pattern is expected after the functionalization of the plasmonic sample, the nanoporous-gold leaf in our case, with a self-assembling monolayer of molecules. As Fig. 4 shows, it is worth noting a clear dependence on functionalization of the resonance term in the near IR range. Resonance position shifts towards lower energy values, i.e. greater wavelengths, and the estimated wavelength shift is about 41.0 ± 4.4 nm (Table 2). This behaviour seems to confirm the plasmonic nature of this absorption contribution and exhibits the high response to the functionalization of the nanoporous substrate.

*3.3 Infrared spectroscopy*

Figure 5a reports the optical constants obtained from the FTIR transmittance and reflectance measurements. As can be seen in figure 5b, these data are well connected (in the spectral region where the two data set are overlapped) with the data obtained from the spectroscopic ellipsometry, thus confirming the quality of the measurements.

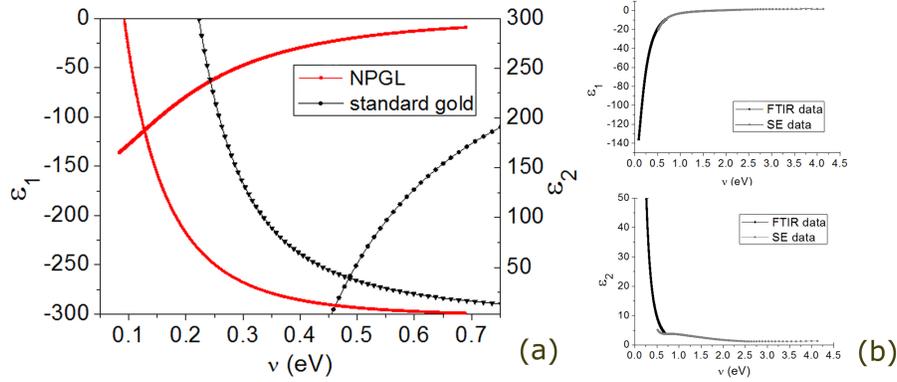

Fig. 5. (a) Comparison between the dielectric permittivity of nanoporous gold leaf (NPGL) obtained from FTIR measurements (red line) and the standard gold: real and imaginary parts. Comparison with standard gold [39] (dark line). (b) Obtained dielectric permittivity in the whole spectral range of analysis.

The obtained optical constants values were used in order to compute the material dispersion curve. A 1D model comprising a thin layer of nanoporous gold (116 nm) was implemented by means of Comsol Multiphysics®, a commercial software based on finite element method analysis. A modal analysis allows computing the expected plasmonic modes and the results obtained with NPGL were compared with the same analysis performed on a self-standing thin layer of standard gold (dielectric constants derived from [41]).

In Fig. 6 we compare the calculated dispersion curves of the optical modes of a 100 nm-thick standard gold layer in air environment (its permittivity being taken from literature [41]) with a similar nanoporous gold layer. Eigenmode analyses of the systems have been carried out by Finite Elements simulations with COMSOL Multiphysics®. As well known [40], the film made of standard gold presents two SPP modes, respectively symmetric (blue dashed line) and antisymmetric (red dashed line) in the transverse field component ($E_z$). The $E_z$ field for the antisymmetric mode is reported in Fig. 6(a). The modes present a maximum effective mode index ($n_{eff}=k/k_0$) close to the surface plasmon resonance frequency, which is located in the visible part of the spectrum. Solid lines in Fig. 6 report the analogous dispersions of the modes calculated for the NPGL gold film.

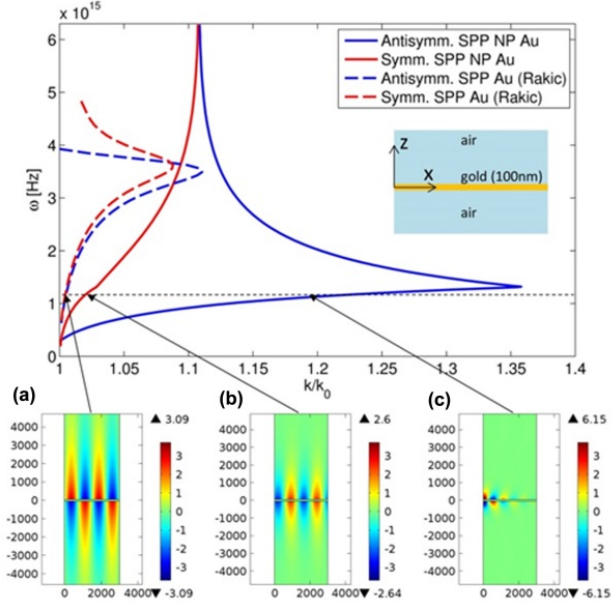

Fig. 6. Modal analysis of plasmonic modes supported by a thin film of bulk gold [40] compared with the ones supported by our NPGL – self-standing configuration. a) Antisymmetric mode in thin gold film immersed in air, b) Symmetric mode in NPGL, c) Antisymmetric mode in NPGL.

As can be seen, there is a strong reduction of the surface plasmon frequency, which falls in the NIR part of the spectrum, as expected. A strong increase in the mode effective index is also observed for the antisymmetric mode, which reaches values as high as 1.35 at the frequency of $1.318 \cdot 10^{15}$Hz ($\lambda=1.43\mu m$). For higher frequency values, the trends of both dispersion curves refer to the excitation of quasi-bound modes [43] and converge to close wavevector values with an effective index around $n_{eff}=1.105$ in the visible range where the material exhibits the behaviour of a dielectric medium with non-null losses. It is worth noting that whereas the antisymmetric-mode curve exhibits anomalous dispersion ($d\omega/dk<0$) as in case of coupled mode in gold thin films [43], the symmetric-mode curve keeps an increasing trend even for frequency values beyond the excitation of surface plasmon modes, in contrast with quasi-bound modes supported by bulk gold films. Inset color plots in Fig.6 exhibits the spatial distribution of the transverse electric field $E_z$ in the film cross-section for three cases of interest at the same frequency: antisymmetric mode in bulk gold film in air (a), symmetric mode in NPGL (b), antisymmetric mode in NPGL (c). Field symmetries confirm mode nomenclature.

## 4. Conclusions

In summary, we presented our work of preparation and characterization of self-standing nanoporous gold leaves. Thin nanoporous gold leaves have been prepared by a reproducible chemical dealloying of a silver-gold alloy film and optically characterized in a broad wavelength range from ultraviolet up to mid infrared by spectroscopic ellipsometry and infrared spectroscopy. Moreover the biosensing capability of the material was investigated by measuring the optical response before and after the functionalization with a self-assembling monolayer of dodecanethiol. Thanks to a greatly enhanced surface-to-volume ratio, nanoporous gold reveals benefits for better reaction efficiency and detection sensitivity. Finally, the material plasmonic properties in the infrared spectral region were investigated looking at the dispersion curve and focusing in particular on the plasmonic properties in this range. A comparison with plasmonic response of a similar bulk gold film highlighted the differences between the two materials.

Thus the use of self-standing nanoporous gold represents a useful and promising approach for the realization of compact plasmonic devices that could be extremely interesting for sensing purposes in a large variety of fields: environmental protection, biotechnology, medical diagnostics, drug screening, food safety and security. The peculiar structure of percolating nanopores and the absence of an obstructing substrate make this material the ideal one for the design and fabrication of chemically stable plasmonic membranes for nano- and micro- fluidic applications in lab-on-chip devices.